\documentclass{emulateapj}
\newcommand{\probg}{\mathrm{P(\textit{galaxy})}}
\newcommand{\probn}{\mathrm{P(\textit{nsng})}}
\newcommand{\probs}{\mathrm{P(\textit{star})}}
\newcommand{\probgn}{\log(p_g/p_n)}
\newcommand{\probgs}{\log(p_g/p_s)}
\newcommand{\probns}{\log(p_n/p_s)}

\slugcomment{Submitted to ApJ}
\shorttitle{SDSS Object Classification}
\shortauthors{Ball et al.}

\begin{document}

\title{Robust Machine Learning Applied to Astronomical Datasets I: Star-Galaxy Classification of the SDSS DR3 Using Decision Trees}
\author{Nicholas M. Ball, Robert J. Brunner, Adam D. Myers}
\affil{Department of Astronomy, MC-221, University of Illinois, 1002 W. Green Street, \\ Urbana, IL 61801, USA \\ National Center for Supercomputing Applications, MC-257, 1205 W. Clark St, \\ Urbana, IL 61801, USA}
\and
\author{David Tcheng}
\affil{National Center for Supercomputing Applications, MC-257, 1205 W. Clark St, \\ Urbana, IL 61801, USA}
\email{nball@astro.uiuc.edu}

\begin{abstract} 
We provide classifications for all 143 million non-repeat photometric objects in the Third Data Release of the Sloan Digital Sky Survey (SDSS) using decision trees trained on 477,068 objects with SDSS spectroscopic data. We demonstrate that these star/galaxy classifications are expected to be reliable for approximately 22 million objects with $r \lesssim 20$. The general machine learning environment Data-to-Knowledge and supercomputing resources enabled extensive investigation of the decision tree parameter space. This work presents the first public release of objects classified in this way for an  entire SDSS data release. The objects are classified as either \textit{galaxy}, \textit{star} or \textit{nsng} (neither star nor galaxy), with an associated probability for each class. To demonstrate how to effectively make use of these classifications, we perform several important tests. First, we detail selection criteria within the probability space defined by the three classes to extract samples of stars and galaxies to a given completeness and efficiency. Second, we investigate the efficacy of the classifications and the effect of extrapolating from the spectroscopic regime by performing blind tests on objects in the SDSS, 2dF Galaxy Redshift and 2dF QSO Redshift (2QZ) surveys for which spectra are available. We find that, for a sample giving a maximal combined completeness and efficiency, the completeness values are 98.9\% and 91.6\% and the corresponding efficiencies are 98.3\% and 93.5\% for galaxies and stars, respectively,  in the SDSS.  We also test our star-galaxy classification by studying the inverse of this sample, finding quasars in the nsng sample with a completeness and efficiency of 88.5\% and 94.5\% from the SDSS and 94.7\% and 87.4\% from the 2QZ. Given the photometric limits of our spectroscopic training data, we effectively begin to extrapolate past our star-galaxy training set at $r \sim 18$. By comparing the number counts of our training sample with the classified sources, however, we find that our efficiencies appear to remain robust to $r \sim 20$.  As a result, we expect our classifications to be accurate for 900,000 galaxies and 6.7 million stars, and remain robust via extrapolation for a total of 8.0 million galaxies and 13.9 million stars. The latter sample should prove useful in constructing follow-up spectroscopic surveys. Characterizing the success of our classifications fainter than $r \sim 20$ will require fainter spectroscopic data.
\end{abstract}

\section{Introduction} \label{Sec: Intro}

It is well-known that classification of objects in astronomical survey data provides a fundamental characterization of the dataset, and forms a vital step in understanding the ensemble properties of the dataset and the properties of individual objects. In this paper, we apply automated machine learning algorithms to classify the 143 million non-repeat objects in the third data release \citep[DR3;][]{abazajian:dr3} of the Sloan Digital Sky Survey (SDSS\footnote{\url{http://www.sdss.org}}; York et al. 2000). The SDSS is used because the combination of the quality and accuracy of the survey CCD photometry, the approximately concurrent spectroscopy performed with the same telescope, and the large number of objects, is ideally suited to the methods we employ in this work. 

As our primary goal is to perform reliable star-galaxy separation, we have used three classes: \textit{galaxy}, \textit{star}, and \textit{nsng} (neither star nor galaxy).  The use of this third class is novel, and is included to not only cleanly separate out those objects that clearly are neither stars nor galaxies (e.g., quasars), but to also improve the extrapolation of our classifications well beyond the  nominal magnitude limits of our current training sample. The latter benefit arises since star-galaxy separation becomes less accurate at fainter magnitudes due to source confusion. In this case, our algorithm can assign sources to the third category when a clear star-galaxy classification is not feasible, thereby minimizing contamination for lower signal-to-noise sources. 

In this paper, the learning algorithm we employ is the axis-parallel decision tree. A decision tree is trained on the subset of data for which spectra are available and is subsequently used to classify the much larger sample for which there is just photometry. This is the first public release of objects classified using this method for the entire SDSS DR3. A natural question is ``Why provide classification for all objects, most of which are near the photometric limits of the survey, where classifications are less robust?'' We have three reasons for classifying the entire SDSS DR3. First, given the speed of applying our classifications, we have found that it is considerably easier to handle large datasets when no artificial cuts are used. With the growth in the number of large astronomical surveys that are accessible from a virtual observatory, we feel this point is becoming increasingly more important, as it greatly simplifies any data federation issues a fellow scientist may need to perform if they wish to use our classifications. Second, classifying all objects allows others with fainter training data to verify the efficacy of our classifications. Finally, since our classifications are probabilistic in nature, by providing the classifications for all objects, we allow others to make their own probability cuts, perhaps using combinations of classes, to perform their own science. For example, our approach simplifies the task of performing follow-up spectroscopy, as a user of our fully classified catalog has an easier task in characterizing any selection effects.

Decision tree learning algorithms have been in existence since the 1970s, see, e.g., \citet{brei:dt} or \citet{quinlan:dt}. They are able to classify objects into discrete or continuous categories, scale well to large datasets, are fairly robust to inequitably-sampled training sets, and can cope with values from individual objects that happen to be bad or irrelevant for the training targets being considered. Axis-parallel decision trees, particularly smaller ones, are also easily understood, providing clear criteria for the splitting of the data set into constituent classes or values. This is in contrast to, for example, artificial neural networks \citep[e.g.,][]{ball:ann}, where the resulting weights are basically a \textit{black box}, or oblique decision trees \citep[e.g.,][]{murthy:odt}, where linear combinations of parameters, or hyper-planes, are used to subdivide the data. Given the size of our dataset, however, our trees are generally rather large, making it difficult to fully interpret the classification rules, but one can in principle follow the splitting of the population. While oblique trees may be less transparant in their operation, they do have the benefit of generally producing smaller trees. Most astronomical results using neural networks and decision trees show that the two methods give comparable results---an example is \citet{bazell:ensembles}, described below.

A number of previous efforts have used decision trees to classify objects in astronomical surveys. \citet{salzberg:dt} use decision trees to separate stars and cosmic ray hits in Hubble Space telescope images and are able to do so with an accuracy of over 95\%. Their algorithm, known as OC1, is an oblique decision tree. \citet{owens:dt} use a version of OC1 to classify galaxy morphologies in the European Southern Observatory--Uppsala surface photometry catalog \citep{lauberts:esouppsala} and show their results to be comparable to those of \citet{storrielombardi:ann}, who use artificial neural networks on the same data. 

\citet{white:first} construct a radio-selected sample of quasars using the VLA-FIRST survey \citep{becker:first} and the optical APM-POSS-I quasar survey of \citet{mcmahon:apmqso}. They show that using decision trees improves on simpler methods and enables quasars to be selected with 85\% efficiency. They also show the completeness and efficiency of the selection as a function of threshold probability from the decision tree that the object is a quasar. \citet{bazell:ensembles} compare Naive Bayes, decision tree and neural network classifiers for morphological galaxy classification and show that the latter two are comparable. 

\citet{suchkov:dt} take advantage of the much improved data, both in quality and quantity of objects, available from the SDSS. They apply the oblique decision tree classifier ClassX, based on OC1, to the SDSS DR2 \citep{abazajian:dr2} to classify objects into star, red star (type M or later), AGN photometric redshift and galaxy photometric redshift bins. Three representative samples, each of $10^5$ photometric objects, one of which extends 2 magnitudes fainter than the spectroscopy, are also investigated. They give a table of percentage correct classifications from which one can calculate completeness and efficiency.

Many studies, too numerous to list here, use other machine learning techniques to classify astronomical survey objects, including star-galaxy separation and quasar identification. These previous works, however, do not provide spectroscopically-trained classifications with probabilities on the scale of the 143 million sources given here for the SDSS DR3. As part of our work, we make publicly available\footnote{\url{http://quasar.astro.uiuc.edu/rml}} the full set of our 143 million object classifications for the SDSS DR3 and probability cuts to select galaxies and stars for a range of completenesses and efficiencies.

The rest of the paper is as follows. In \S \ref{Sec: Data} we describe the datasets used in the construction and testing of the catalog. \S \ref{Sec: Algorithms} describes the decision trees, their optimization using supercomputing resources, and the probability cuts to create subsamples of particular objects. \S \ref{Sec: Results} then describes the optimal learning parameters found and the resulting datasets. In \S \ref{Sec: Discussion} we discuss the efficacy of our star-galaxy separation, the extrapolation of our results to fainter magnitude, and new projects we have undertaken to extend this work. We conclude in \S \ref{Sec: Conclusions}.

\section{Data} \label{Sec: Data}

The SDSS is a project to map $\pi$ steradians of the northern Galactic cap in five bands ($u$, $g$, $r$, $i$ and $z$) from 3500--8900 \AA. The final survey will provide photometry for approximately $2.5 \times 10^{8}$ objects, of which a multifiber spectrograph will provide spectra and redshifts for around one million. The spectroscopic targeting pipeline selects targets for spectroscopy from the imaging \citep{eisenstein:sdsslrg, richards:sdssqso, strauss:mainsample}, and a tiling algorithm \citep{blanton:sdsstg} optimally assigns spectroscopic fibers to the selected targets. Further details of the survey are given in \citet{york:sdss} and in \citet{abazajian:dr3} for DR3.

The datasets we use in this work consist of training, testing\footnote{Note that some papers in the astronomy literature use the term `validation set' to refer to what is here described, in the usual machine-learning sense, as the testing set. The validation set can be used, for example, in the training of artificial neural networks to determine when to stop training in order to avoid overfitting.}, blind testing, and application sets. The training and testing sets are drawn from objects in the \texttt{specObj} view of the DR3 Catalog Archive Server (CAS) public database\footnote{\url{http://cas.sdss.org/astrodr3/en}}. The blind testing sets are from the SDSS DR3, 2dF Galaxy Redshift Survey \citep[2dFGRS;][]{colless:2dffinal} final data release\footnote{\url{http://www.mso.anu.edu.au/2dFGRS}} and the 2dF QSO Redshift Survey \citep[2QZ;][]{croom:qsolf} final data release\footnote{\url{http://www.2dfquasar.org}}. The application set is a locally hosted version of the \texttt{photoPrimary} view of the DR3 CAS. As with \texttt{specObj}, this is the view which contains the primary SDSS observations with no duplicates. The raw \texttt{specObj} data consists of 477,081 objects and the \texttt{photoPrimary} data 142,705,734\footnote{A previous version of the DR3 CAS had a slightly different total number, 142,759,806, traceable to objects in run 2248.}.

The training, testing and application data consist of SDSS photometric and spectroscopic attributes available in the \texttt{photoPrimary} and \texttt{specObj} tables. We perform no additional image or spectral reduction. Since we classify all sources in the SDSS DR3, this allows a user of our classifications to apply their own sample cuts based on SDSS flags or attribute errors.  The photometric attributes used for each object are the colors $u-g$, $g-r$, $r-i$, and $i-z$, corrected for galactic reddening via the maps of \citet{schlegel:dustmaps}. The colors in each of the four magnitude types PSF, fiber, Petrosian and model are used. The use of the four magnitude types allows star-galaxy separation to be, to some extent, built-in. The light profiles of galaxies and stars are very different, which can be quantified by measuring the object flux in different-sized apertures. This fact is often used to perform star-galaxy separation,  including within the SDSS photometric pipeline where the difference between  $psfMag$ and $cModelMag$ is used (see \S~\ref{Sec: Algorithms} for more information). The spectroscopic attribute is the target object type and is given by the \texttt{specClass} value in \texttt{specObj} assigned by the SDSS spectroscopic pipeline.

The \texttt{specClass} is discrete and takes the values \texttt{unknown}, \texttt{star}, \texttt{galaxy}, \texttt{qso}, \texttt{qso\_hiz} or \texttt{star\_late}. We construct our three training samples: galaxies, stars, and nsng, from these classes according to the mappings shown in Table \ref{table: mappings}\footnote{There are also the classes \texttt{sky} and \texttt{gal\_em} (emission-line galaxy) in the DR3 database, but there are no examples of the former in \texttt{specObj} and the latter is a placeholder.}. Resulting numbers of each are 361,502 galaxies, 62,333 stars and 53,246 nsng, giving a total of 477,081 objects available for the training and testing process. Given the spectroscopic data available in the SDSS, the nsng training sample will predominantly consist of quasars. Decision trees, however, effectively utilize all of their training data. Thus, at fainter magnitudes, we can expect the nsng sample to contain sources that are not just quasars, but unknowns in conjunction with the \texttt{unknown} category in the \texttt{specClass}. 

On the other hand, decision trees are also robust to outlying values in their training data. For the classification training we simply apply the very broad cuts $-40 < {\mathrm{color}} < 40$ to remove clearly unphysical values from the data such as -9999. These are applied to each of the colors and resulted in the removal of 12 galaxies and 1 quasar, leaving 477,068 for training and testing.

For the decision tree optimization, sets of objects were randomly drawn from the remaining \texttt{specObj} set and used in training, with the rest for testing. Thus the testing set, in a strict sense, is not independent of the training set. However, the large numbers of objects available in the training data means that the random subsamples drawn, except perhaps the smallest ones at a level of 10\% or less, are large enough to still be truly representative of the training data and should thus be effectively independent.

The performance results quoted in \S \ref{Sec: Results} are, nevertheless, from a tree trained \textit{and} tested on 80\% of the training data and applied to the other 20\%. The 80\% is subdivided as above into non-overlapping training and testing sets, also in the ratio 80:20. Given the hypothesized easily-adequate size of the training set, this pseudo-blind test should be very similar to the test results during the optimization.

For the final application to the \texttt{photoPrimary} data, the tree was trained on all of the training data to maximize the available information. Thus the results quoted in \S \ref{Sec: Results} may be slightly pessimistic, but again they are likely to be very similar to the performance of the tree on \texttt{photoPrimary}, within the spectroscopic regime. For the vast majority of objects in the \texttt{photoPrimary} table, no spectroscopic information is available. In addition, we do not restrict the application dataset,  as each object is classified independently. Therefore, specific cuts, either on SDSS flags, attribute errors, or the attributes themselves can be later applied by a user as appropriate to create a particular sample. For example, one might only want `best' photometry, as indicated by the object flags, for comparing object populations, or one might want to apply a color cut to enhance the utility of the provided classifications.

The blind testing sets are drawn from the 2dFGRS and the 2QZ. These test the performance of the decision tree on non-SDSS data in a parameter space which extends beyond the SDSS DR3 spectroscopic regime. The 2dFGRS was used to test the best tree's galaxy classifications and the 2QZ to test its star and nsng classifications. The surveys were matched with the SDSS using object position with a tolerance of 2 arcseconds and their spectroscopic classifications were taken as the target types. Of the 2dFGRS galaxies, 50,191 were matches. For the 2QZ the results were compared to the objects assigned the category \textit{11} (best classification and redshift; see \citealt{croom:qsolf}) objects (8,739 quasars, 5,193 stars) and those assigned \textit{12}, \textit{21} or \textit{22} (second-best; 9,273 quasars).

\section{Algorithms} \label{Sec: Algorithms}

Star-galaxy separation in the SDSS is currently performed within the SDSS photometric pipeline \citep{lupton:photo}. An object is classified as extended, and hence a galaxy, if ${\mathrm\texttt{psfMag}} - {\mathrm\texttt{cmodelMag}} > 0.145$. Here \texttt{psfMag} is the point-spread-function magnitude and \texttt{cmodelMag} is a combination of the best fitting de Vaucouleurs and exponential profiles. The separation is done for each band individually, and for all five bands combined. \texttt{psfMag} is described further in \citet{stoughton:edr} and \texttt{cmodelMag} in \citet{abazajian:dr2}. Additional star-galaxy classifications have been performed on SDSS data, e.g., \citet{scranton:sdsssys} perform a Bayesian star-galaxy separation via differences between the $r$ band \texttt{psfMag} and \texttt{modelMag}.

In our approach, we use an axis-parallel decision tree to assign a probability that a source belongs to each of three classes. These probabilities always sum to one and reflect the relative degree of certainty about an object's type. We provide cuts in these probabilities that can be applied to generate catalogs to a required completeness and efficiency of either galaxies, stars, or neither galaxies nor stars (i.e., nsng). In general, the cuts can be a function of the probabilities, and these functions can have both minimum and maximum values.

We define both completeness and efficiency as functions of the classification probability. For a given sample, therefore, the \textit{completeness} is the fraction of all catalog sources of a given type found at a specific probability threshold. Likewise, \textit{efficiency} is the fraction of all catalog sources that are correctly classified to a specific probability threshold. In previous astronomical machine learning results, the efficiency of a sample has also been called the \textit{reliability} or \textit{purity}, and $1.0 - $\textit{efficiency} has been called the \textit{contamination}. The quantities are also known \citep[e.g.,][]{witten:datamining} as the \textit{recall} and \textit{precision}.

As well as a maximal combination of completeness and efficiency, one may also desire maximal values of these measures separately. An example of the former is to include interesting or unusual objects of a certain class, for example, candidate objects for gravitational lensing. An example of the latter is applications such as the 2-point correlation function of extragalactic objects, where one wants to minimize contamination by non-cosmological objects such as Galactic stars \citep[e.g.,][]{myers:qsoevoln}.

Our decision tree methods are applied within the framework of the Data-to-Knowledge Toolkit\footnote{\url{http://alg.ncsa.uiuc.edu/do/tools/d2k}} \citep{welge:d2k}, developed and maintained by the Automated Learning Group at the National Center for Supercomputing Applications (NCSA). This is a Java-based application which allows numerous modules, each of which performs a single data processing task, to be interconnected in a variety of ways. The framework allows the easy addition of further data to the training and testing process and the application of the trained decision trees to the entire photometric catalog. The trees are implemented on the Xeon Linux Cluster \textit{tungsten} at NCSA. This nationally allocated supercomputing system is composed of 1280 compute nodes. Each node has available 2 GB of memory and a peak double-precision performance of 6.4 Gflops.

Our use of the \textit{tungsten} supercomputing system was facilitated by a large, peer-reviewed, national allocation submitted by the Laboratory for Cosmological Data Mining at NCSA, for which RJB is the principal investigator. Execution tasks were submitted to \textit{tungsten} via a shell script that created and tested one or more decision trees whose testable parameters were specified at run time. One or more decision trees would be created and tested on a single \textit{tungsten} node, and when completed the test results for each decision tree was returned for subsequent analysis. As this process was embarrassingly parallel, we submitted multiple execution tasks simultaneously from this single shell script, with a maximum of 240 submitted at any one time. As discussed in the next section, we analyzed the results of over seven thousand decision trees, which spanned a range of different decision tree parameters in selecting our optimal parameter values. 

\subsection{Decision Trees} \label{Subsec: Algorithms - DTs}

Decision trees consist first of a root node in which the parameters describing the objects in the training set population are input, along with the classifications. The tree is usually considered upside-down, with the root node at the top. Here there are three classes, so for each object we have \begin{equation} {\mathrm{input}} = ([f_1, f_2, f_3, \ldots, f_{16}],[o_g,o_n,o_s]) \end{equation} for the 16 features $f$ (four colors in four magnitude types) and the three output classes $o$. The outputs are [1,0,0], [0,1,0] and [0,0,1] for galaxy, nsng, and star, respectively. The outputs are in the form \begin{equation} {\mathrm{output}} = [p(o_g), p(o_n), p(o_s)], \end{equation} and the predicted probabilities of class membership always sum to one.

A node population is split into population groups that are assigned to child nodes using the criterion that minimizes the classification error. This is a measure of how good the classification is when the tree that would result from the split is run on the test set. The process is repeated iteratively, resulting in a number of layers of nodes that form a tree. The iteration stops when either all nodes reach the minimum allowed population of objects in a node (the minimum decomposition population; MDP), the maximum number of nodes between the termination node and the root node (the maximum tree depth; MTD), or the population split no longer decreases the classification error by a minimum set amount (the minimum error reduction; MER). The nodes from which no further nodes branch are the leaf nodes.

The computational complexity of the algorithm is $O(n m d)$, where $n$ is the number of examples, $m$ is the number of features, and $d$ is the depth of the tree. Here $n$ is $O(10^5)$, whereas $m$ and $d$ are $O(10)$, giving a product of $O(10^7)$.

The split is tested for each input feature. We use axis-parallel splits, in which, following \citet{owens:dt}, \begin{equation} a_{i}X > k, \end{equation}  where $a_{i}$ is the $i$th attribute for example $X$, and $k$ is a value to be tested. An alternative, which we have not utilized in this work, is the oblique split \begin{equation} \sum_{i=1}^{d} a_{i} X_{i} + a_{d+1} > 0 \end{equation} for $d$ attributes. This allows hyperplanes at arbitrary angles in the parameter space. Here we allow the split point to be either the midpoint, the mean, or the median of the values of the parameters for the node population, with the best of these chosen for each individual node. With the removal of extreme non-physical outliers such as -9999 from the training data, selecting the optimal statistic from those measured at each node splitting provides better results than consistently selecting the same statistic.

The classification error we used is the variance \begin{equation} \sum_{i=1}^{3} (A_i - P_i)^2, \end{equation} where $i$ is the output feature index, $A$ is the actual output feature value (0 or 1), and $P$ is the probability prediction made by the decision tree. The data is modeled using the mean of all the output vectors, and the result is interpreted as a probability over class space.

We used $n$-fold bagging (bootstrap-aggregating) in addition to the optimized tree and node parameters. In this procedure, the testing set is a randomly sampled fraction (bagging fraction) of the original testing set, and the procedure is repeated $n$ times, creating $n$ decision trees. The results are then voted on, using simple majority voting. Similarly, the testing set itself is randomly subsampled from the training set, a procedure known as cross-validation. We perform 10-fold bagging, but use 1-fold cross-validation, as only the former improved the results.

The parameters of the decision tree, and additional methods such as bagging, can have a substantial effect on the quality of the results produced. The massively parallel supercomputing resources we used in this work allowed extensive investigation of this parameter space, including: 1) the minimum decomposition population, 2) the maximum tree depth, 3) the minimum error reduction, 4) the method of splitting the population at a node, either half-and-half, midpoint, mean or median, 5) the ratio of sizes of the training set to the testing set, 6) the bagging fraction, and 7) the random seed used in selecting the subsamples for the testing set and the bagging. In addition, the construction of thousands of trees and the use of advanced visualization allowed us to investigate the interaction of these parameters.

\subsection{Probabilities} \label{Subsec: Algorithms - Probabilities}

Once we have constructed an optimal decision tree, given the available training data, we can characterize the best probability thresholds to construct samples from further data to the desired level of completeness or efficiency. We investigated the basic probabilities $\probg$, $\probn$ and $\probs$ (hereafter $p_g$, $p_n$, and $p_s$) and the ratios \begin{equation} f(p) = \probgs, \probgn, \quad {\mathrm{and}} \quad \probns \end{equation}  for each target type galaxy, star and nsng for binned minima and/or maxima, using the 80:20 pseudo-blind test described in \S \ref{Sec: Data}. 

Small numbers of objects in the leaf nodes resulted in discretization in the output probabilities from the decision tree. Therefore, care was taken when comparing the resulting mixture of exact and floating point values to bin edges to ensure that the counts were correct and not offset due to finite floating point precision. The accuracy was achieved by using a Decimal datatype, which stores numbers as exact values to a specified number of decimal places.

\section{Results} \label{Sec: Results}

\subsection{Decision Tree Parameter Optimization} \label{Subsec: Results - DT optimization}

As described in \S \ref{Subsec: Algorithms - DTs}, a decision tree has a number of parameters that can be adjusted. The product of these parameters defines an enormous parameter space of possible trees that must be explored to identify what we expect to be the optimal decision tree. In addition, this parameter space can potentially include many local minima of the classification error, further complicating this process. Given the size of the parameter space and the computational time required to construct and test an individual decision tree, we utilized the NCSA \textit{tungsten} cluster (described in more detail in \S \ref{Sec: Algorithms}) to perform an extensive exploration of this parameter space. All decision trees were built from the same training set with the 16 input parameters described in \S \ref{Sec: Data}, and the full parameter space we explored is described in Table \ref{table: dt params}. 

\begin{figure}
\figurenum{1}
\epsscale{0.5}
\plotone{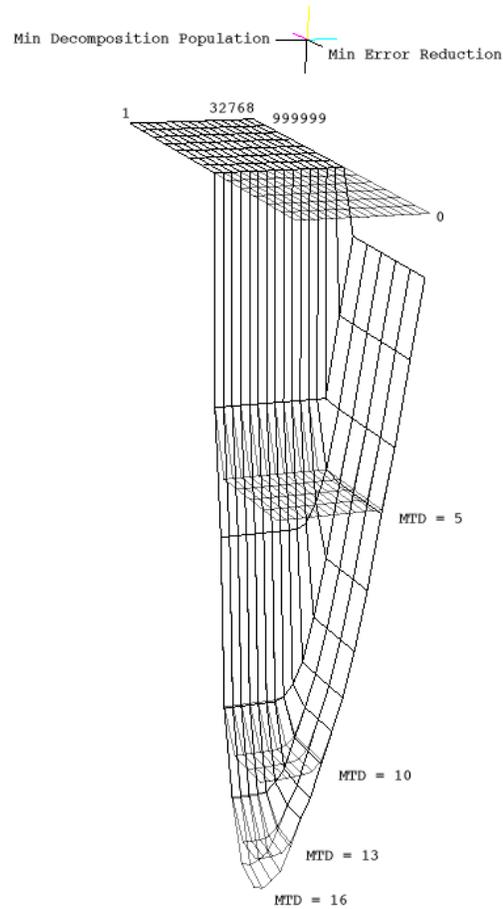}
\caption{Partiview visualization of the decision tree classification error from training as a function of the minimum decomposition population (MDP) and minimum error reduction (MER) for the maximum tree depths (MTDs) shown (including 1, the front extension of the plane at the highest classification error). Each vertex of the mesh represents the result from a decision tree. The tree depth was memory-limited to 16 but as the depths not shown follow the pattern of those shown it is clear that there will not be significant improvement. The error ranges from 24\% for a depth of 1 to 3\% for a depth of 16. The best MDP is 2 and the best MER is 0. Their values are, respectively, $2^a$ and $10^b$, where $0 \leq a \leq 15$, $-6 \leq b \leq 6$, and MER $ = 0$. An MER below zero gives the same result as the single layer tree. These and the other decision tree values used are given in Table \ref{table: dt params}. \label{fig: dt params}}
\end{figure}

First, we studied the interaction of the minimum node decomposition population (MDP), the maximum tree depth (MTD) and the minimum error reduction (MER) for splitting the population. The values investigated were $\mathrm{MDP} = 2^a$, where $0 \leq a \leq 15$, $1 \leq \mathrm{MTD} \leq 20$, and $\mathrm{MER} = 10^b$, where $-6 \leq b \leq 6$, and $\mathrm{MER} = 0$. Negative MERs do not split the population. This gave 4,480 combinations, and for each one a decision tree was generated. Figure \ref{fig: dt params} shows a Partiview~\citep{levy:parti} visualization of the results. The best tree had an MDP of 2 and an MER of 0. MTD was limited to 16 by the combination of the size of the training set and the memory available on tungsten but, as shown in Figure \ref{fig: dt params}, a deeper tree is unlikely to give a substantial improvement.

We next varied the statistic used to split the population at the nodes: halving the population, using the midpoint of each input, the mean of each input, and the median. The 16 combinations give very similar results, with the exception of those which allow population halving, which are significantly worse. As a result, we allowed the algorithm to select, when splitting each node, the optimal statistic from the midpoint, mean, and median split values.

In all previous measurements, we had set the level of bagging used when building the decision tree at 20-fold and assumed a ratio of the fraction of the training set used in the bagging to the fraction of the sample used in training of 80:20 (i.e., 80\% for training to 20\% for testing). We conducted a number of tests to verify these assumptions. When varying the level of bagging, we found that the classification error is substantially reduced when the bagging level is increased from 1- to 10-fold, and continues toward an asymptotic value slightly below the value seen for the 50-fold case. As with our tests involving the maximum tree depth, the number of models was limited by the available memory on each node of the tungsten cluster, although given the asymptotic tendency, we do not expect that the results would change substantially for larger levels of bagging.

\begin{figure}
\figurenum{2}
\plotone{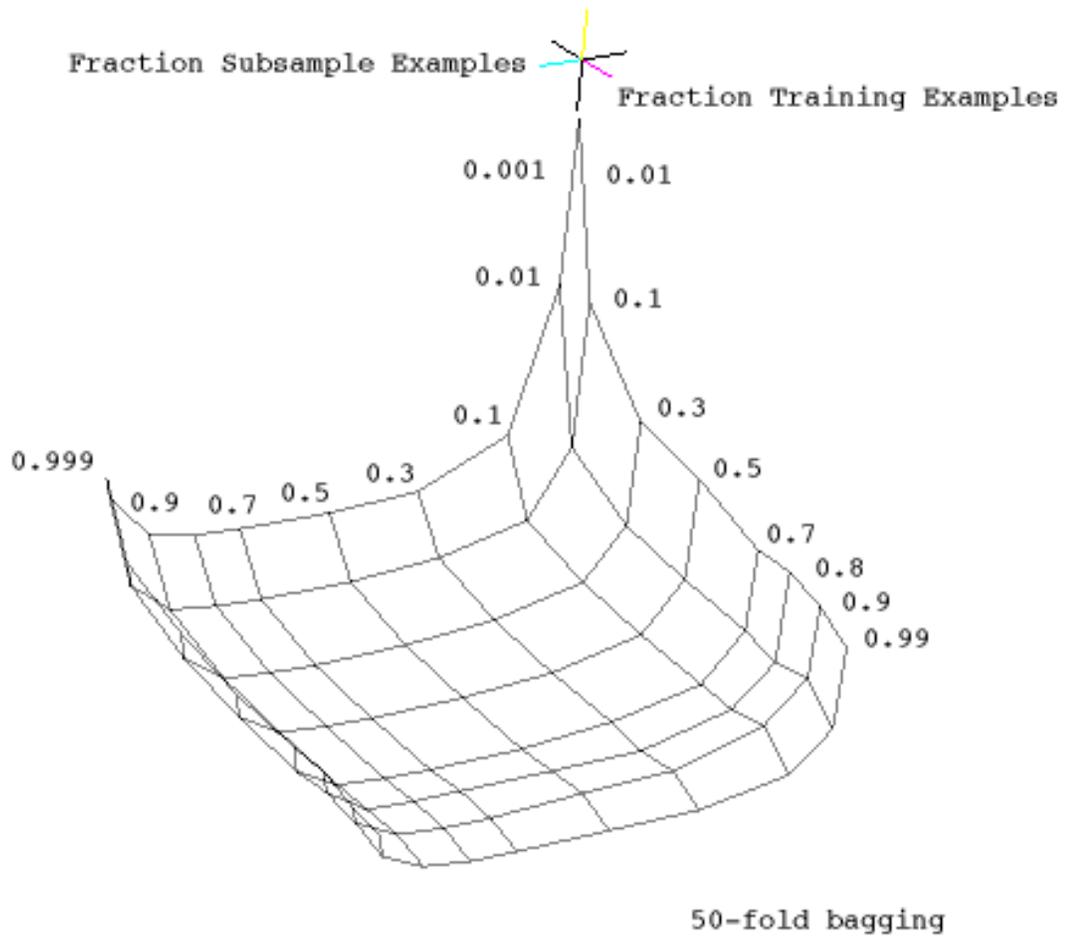}
\caption{As Figure \ref{fig: dt params} but showing the bagging fraction of subsample examples (FSE) and the training set fraction (FTE) for 50-fold bagging. Here one can see the broad flat base corresponding to the many trees which approximately achieve the minimum classification error of 3\%. Numerous other plots show this broad base, providing evidence that further improvement in performance can only come from improved training data. The best FSE and FTE values are both 0.8. The 0.99 FTE is lower due to a very small test set and is not used. The high FSEs have worse errors. Results for 10- and 20-fold bagging are similar but with slightly higher classification error and more pronounced change at high FSE and FTE. \label{fig: bagging}}
\end{figure}

We cross-tested the training to testing ratios with the n-fold bagging test, finding the best results were obtained using a 50-fold model with an 80\% training subsample, as shown in Figure \ref{fig: bagging}. Given memory limitations, the actual trees used in our other parameter tests were always limited to a 20-fold bagging. When performing our final classifications, however, we used the more accurate 50-fold bagging. In all subsequent tests we set the training to testing fractions to 80:20. We also note that the bagging prevents an overfitting of the training set: for low values of bagging, such as 1-fold, the classification error has a minimum at a particular MTD, and increases substantially for deeper trees.

\begin{figure}
\figurenum{3}
\plotone{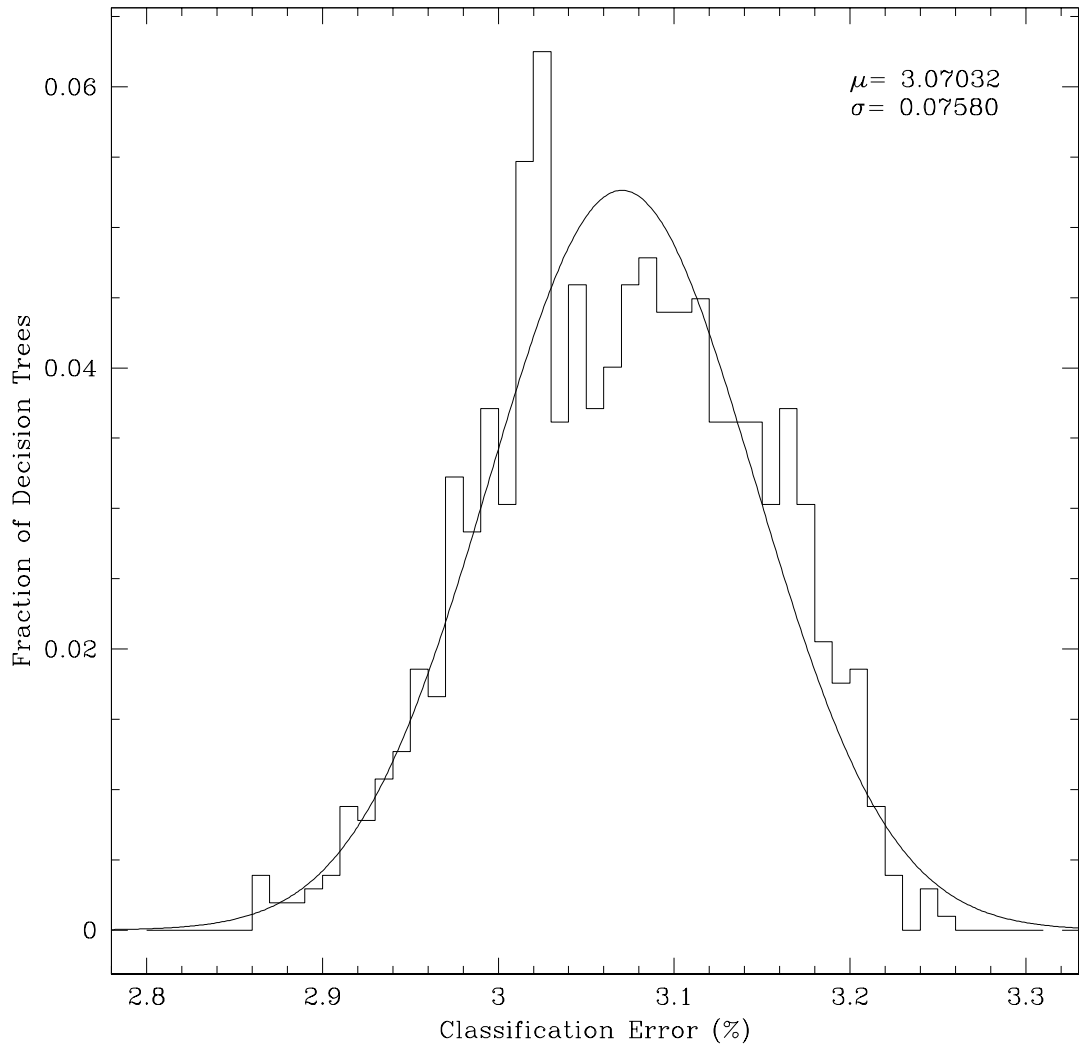}
\caption{Normalized histogram of the effect of random seed on the decision tree performance, showing the distribution of the percentage classification error for instances of different random seeds in the selection of subsamples for training and for bagging. 32 seeds were chosen for each, and the tree for each combination computed. The mean is 3.07\%, the standard deviation is 0.08, and a Gaussian with these values is overplotted, normalized to the same amplitude as the histogram. \label{fig: random seed}}
\end{figure}

Finally, the effect of the random seed in choosing a same-sized training set was investigated and is shown in Figure \ref{fig: random seed}. The $1\sigma$ variation in the classification error is approximately 0.1\%. Given this variation, we quote the best classification error achieved as being $(3.07 \pm 0.08)\%$. This is a robust result as the classification error shows a broad, approximately flat, minimum in several areas of the parameter space. See, for example, Figure \ref{fig: bagging}, where the flat area corresponds to approximately 3\% classification error.

Although our exploration of the parameter space was extensive, the number of \textit{possible} trees in the space quickly becomes very large with the variation of all parameters, so it is always possible that a better tree exists than the one we used. Although given the widely observed 3\% minimum it is unlikely to be \textit{significantly} better, especially for the training set we used.

\subsection{Testing Star-Galaxy Separation and Probability Cuts} \label{Subsec: Results - Testing}

Once the optimal decision tree parameters were quantified, we constructed and subsequently tested this decision tree using unseen data to give a more realistic idea of the quality of our classifications. We achieved this result by building the tree with the previously described parameters, but training on 64\% of the DR3 data and testing on 16\% (i.e., maintaining the optimal 80:20 ratio). We tested this decision tree in a pseudo-blind fashion by using the remaining and unseen 20\% of the DR3 data, which consisted of 95,413 sources. From this classification test, we compared the assigned types galaxy, star and nsng to the true types, in terms of completeness and efficiency as a function of the three probabilities $p_g$, $p_s$, and $p_n$. 

\begin{figure}
\figurenum{4}
\plotone{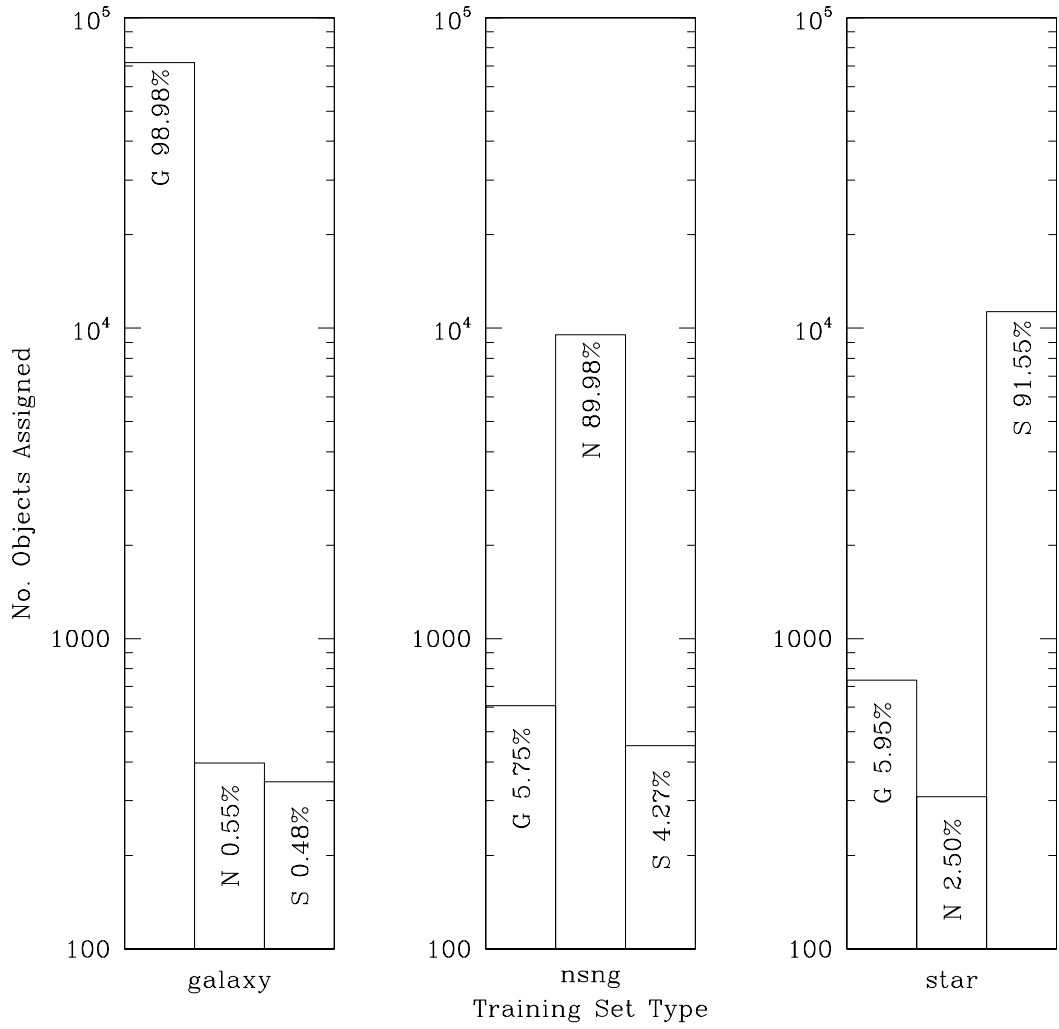}
\caption{Confusion matrix for the overall assignment of the types galaxy, nsng and star for the SDSS DR3 blind testing set in the spectroscopic regime. The three columns in each panel show the numbers of galaxies, nsngs and stars assigned and hence that the decision trees successfully assign the types to the objects. The values are tabulated in Table \ref{table: confusion matrix}. The vertical axes are logarithmic to clarify the significant differences between the number of successfully and unsuccessfully classified objects. The percentages in each panel add to 100\%. \label{fig: confusion matrix}}
\end{figure}

\begin{figure}
\figurenum{5}
\plotone{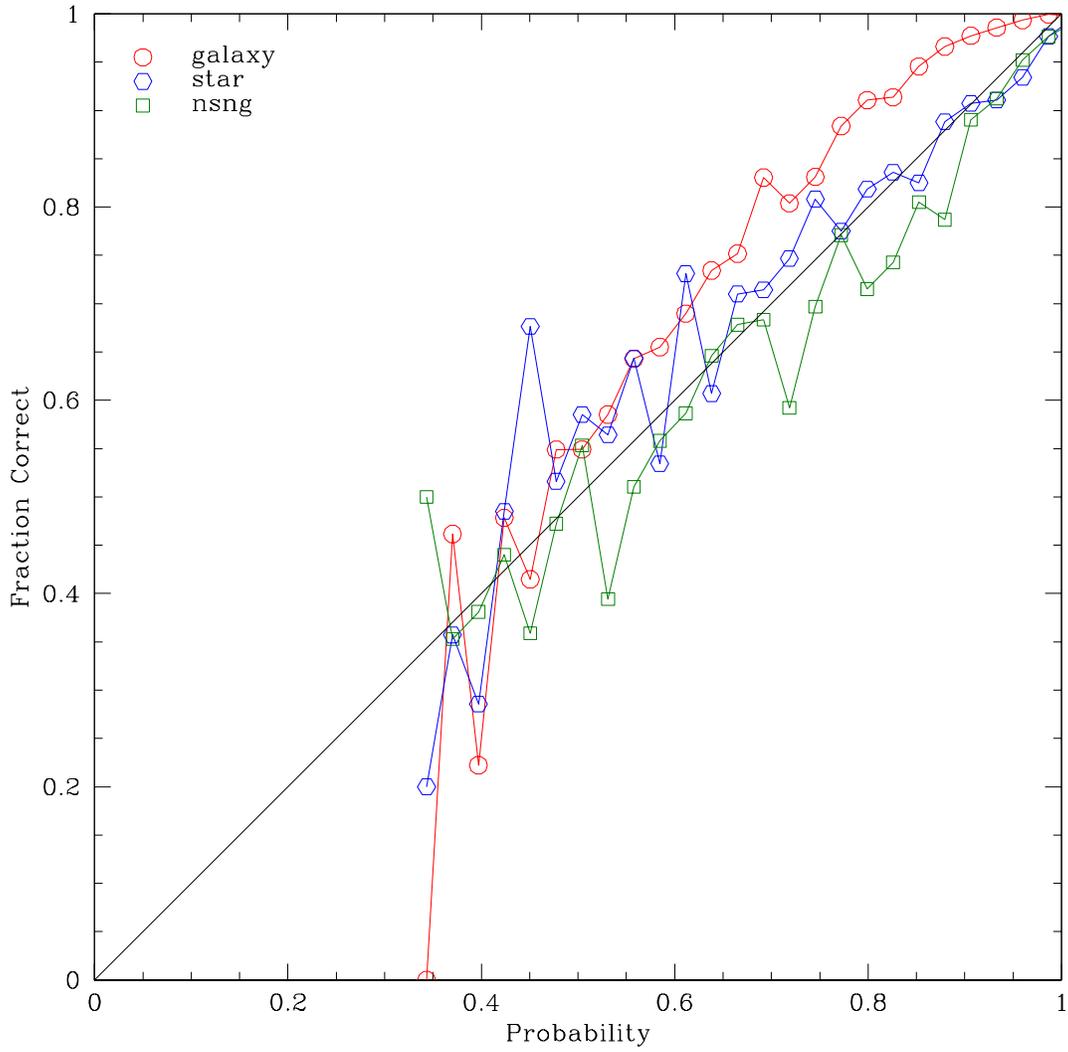}
\caption{True fraction of galaxies, nsngs and stars compared to the probability assigned by the decision tree that the object is of that type. The solid line shows the ideal result of equal values for each object. \label{fig: true probability}}
\end{figure}

Overall, this experiment demonstrated that our axis parallel decision tree is very successful at classifying objects into these three types, and, as a result, at star-galaxy separation. The vast majority of objects are classified correctly and given high probabilities for the appropriate type. The confusion matrix is shown in Figure \ref{fig: confusion matrix} and tabulated in Table \ref{table: confusion matrix}. Figure \ref{fig: true probability} compares the true fraction of correct classifications as a function of assigned probability from the tree, demonstrating that these are approximately correct. We note that there is some discretization in the probabilities due to the potentially small number of objects in a leaf node. As can be seen in Figure \ref{fig: dt params}, however, increasing the minimum decomposition population to the point at which the probabilities would not be discretized would worsen the decision tree performance.

In general the function for generating subsamples according to probability is 
\begin{equation} 
f(p) = f(p_g, p_n, p_s). 
\end{equation}  
In all subsequent tests, we analyze a subsample in terms of true positives (TP), false positives (FP), true negatives (TN) and false negatives (FN) for each of the three target types separately. We define the completeness, $c$, as the number of the target type that are included in the $f(p)$ cut compared to the total number of objects of that target type in the whole sample: 
\begin{equation} 
c = \frac{\mathrm{TP}_i}{\mathrm{TP}_{all} + {\mathrm{FN}_{all}}}, 
\end{equation} 
where $i$ indicates the objects to the probability threshold and $all$ indicates all the objects in the test set. For example, in our pseudo-blind test set this would correspond to all 95,413 sources. The efficiency, $e$, is the fraction assigned a particular type that are genuinely of that type: 
\begin{equation} 
e = \frac{\mathrm{TP}_i}{{\mathrm{TP}}_i + {\mathrm{FP}}_i}. 
\end{equation} 
The completeness varies between 0 and 1 and in general increases at the expense of efficiency. An object can be TP or FN depending on the probability threshold. 

We define the best sample for a specific test to be that which maximizes the metric: 
\begin{equation} 
{\mathrm{best}} = {\mathrm{max}}[\sum (c^2 + e^2)^{1/2}]. 
\end{equation}  
All the ratios, but not the simple probabilities, exclude objects for which any of the three probabilities are zero. In our pseudo-blind test, 18,989 of the 95,413 sources, or roughly 20\%, were affected in this manner. We used the logarithm of a specific probability ratio, given the large dynamic range of the ratios of the the three probabilities. In all cases the classification is taken to be the object type assigned the highest probability by our decision tree. If the highest probability is less than 50\% it is possible that two types are assigned an equal probability, which could introduce a systematic effect in our classification. We do not expect this to be problematic in practice, however, as this effect is small, only affecting one object in our pseudo-blind testing, and excluding these sources is easy after classification has been performed.

To determine the best probability function, i.e., $f(p)$, and associated probability values to use when assigning classifications, we computed a number of different functions involving the three probabilities assigned by our decision tree. Table~\ref{table: f(p) cuts} provides a compilation of these different functions and a characterization of their distribution sampled across one hundred bins. From the values in Table~\ref{table: f(p) cuts}, we find that by using the simple minima in the three class probabilities we obtain the most accurate classification samples in our pseudo-blind testing. These probability cuts are $p_g \geq 0.49$, which results in 98.9\% completeness and 98.3\% efficiency for galaxies, $p_n \geq 0.54$, which results in 88.5\% completeness and 94.5\% efficiency for nsng, and $p_s \geq 0.46$, which results in 91.6\% completeness and 93.5\% efficiency for stars. We find that a number of probability cuts near fifty percent produce similar classification accuracies. This is not surprising, as fifty percent is an ideal threshold, as objects below this percentage are likely to be of a different class since they can have a higher probability of being one of the other two types. 

\begin{figure}
\figurenum{6}
\plotone{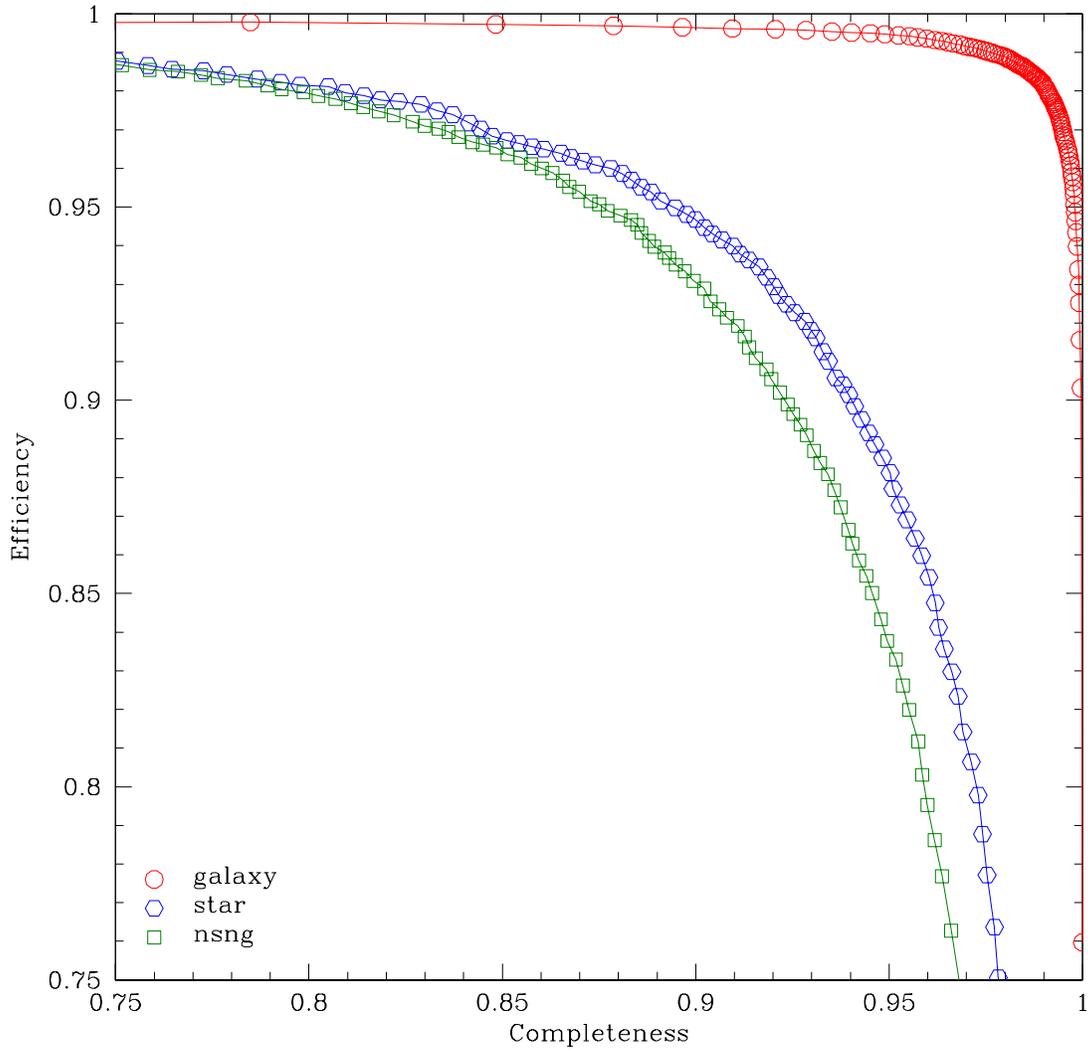}
\caption{Completeness and efficiency for galaxies as a function of $\probg$, $\probn$ and $\probs$ for the 80:20 blind testing set. The optimal samples quoted in \S \ref{Subsec: Results - Testing} correspond to the points closest to the upper right of the plot for each object type. The curves extend outside the axes shown. \label{fig: com eff}}
\end{figure}

In some cases, ratios of probabilities are able to select better samples when a maximum value is also used as opposed to only using a minimum value. For example, for the probability ratio $f(p) = \probgs$, the sample is dominated by stars at the low end and galaxies at the high end. So the nsng are only selected by requiring a minimum and a maximum $f(p)$. In the end, however, none of the ratio cuts we have studied provided a better combination of completeness and efficiency than the simple probability cuts. We interpret this result as further evidence that our decision tree is successful, especially within the regime of spectroscopically confirmed objects used for training (i.e., the interpolation regime). The full completeness-efficiency curves, as a function of $p_g$, $p_n$, and $p_s$, are shown in Figure \ref{fig: com eff}.

\subsection{Application to the SDSS Photometric Catalog} \label{Subsec: Results - photoPrimary}

We constructed our final decision tree from the full (i.e., 100\%), spectroscopic training sample from the SDSS DR3. By using the data-streaming modules available within the D2K environment, we applied this decision tree to the full SDSS DR3 \texttt{photoPrimary} catalog of 142,705,734 sources. With this paper, we publicly release the classification probabilities for every one of these sources. Following the classification guidelines discussed in the previous section, within the magnitude range $0 \leq r \leq 40$, this sample consists of 38,022,597 galaxies, 57,369,754 stars and 47,026,955 nsng. The sum of these subsample sizes do not match the size of the full sample because of (a) the magnitude restriction, which removes 589 objects, and (b) source classifications that are ambiguous due to equal-highest probabilities for multiple object types, which removes 285,839 (0.2\%) objects.

One of the most important issues with any supervised classification effort is the application of an algorithm trained on one sample of data to a different set of data. This concern is generally framed in terms of interpolation versus extrapolation, and in our case results from the application of our decision tree that is trained on sources from the SDSS spectroscopic sample being applied to the SDSS photometric sample, which is, in some cases, considerably fainter than our training set. Given the strong dependence on the number counts of galaxies and stars with apparent magnitude, this affects the vast majority of the sources we have classified. Without deeper training data, however, we feel that the best approach is to classify all sources. This approach allows anyone using our classification catalog to create a well-defined sample for further study or follow-up spectroscopy that does not have any selection effects artificially imposed due to our classification algorithm.

While we cannot guarantee the accuracy of our classifications across the full SDSS photometric sample, we can provide some guidance on the accuracy of specific sample restrictions. Historically, one of the most powerful techniques for characterizing classifications in the absence of spectroscopic identifications is by analyzing the differential number of sources as a function of apparent magnitude (i.e., number count plots). This technique is simple and straightforward; by comparing the number count distribution of training sources with the number count distribution of classified sources, we gain a statistical estimate of the apparent magnitude limit at which we can no longer expect our extrapolated classification rules to remain robust. 

\begin{figure}
\figurenum{7}
\plotone{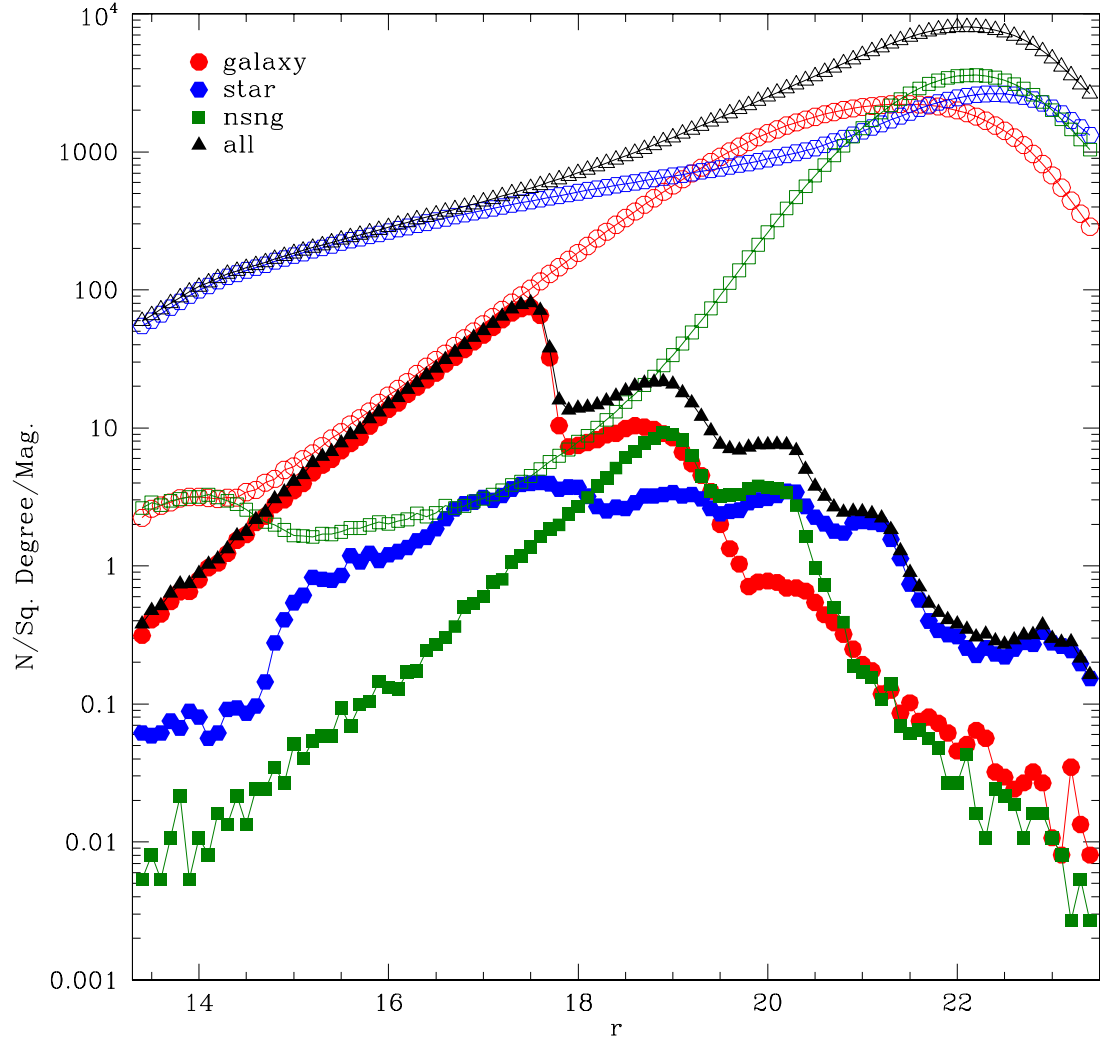}
\caption{Number counts of the objects as a function of extinction-corrected $r$ band model magnitudes, subdivided by object type galaxy, nsng, or star. The lower set of four curves (closed symbols) shows the target types of objects in the training set. The upper set (open symbols) shows the full counts of the types assigned by the decision tree to the SDSS DR3. A small number of objects lie outside the axes shown. The highest three peaks in the galaxy and nsng training set curves correspond to the cuts in the spectroscopy for galaxies, quasars and high-redshift quasars, at magnitudes $r = 17.77$, $r \sim 19$, and $r \sim 20.5$ respectively. The extrapolation in magnitude to the full DR3 is clear. \label{fig: mag counts}}
\end{figure}

We present, therefore, in Figure~\ref{fig: mag counts}, the number counts for both the training set and for the classified \texttt{photoPrimary} catalog for our three different classifications: galaxy, nsng, and star. The SDSS spectroscopy is limited to $r < 17.77$ for galaxies \citep{strauss:mainsample}, which can clearly be seen in the galaxy training set number counts. Similarly, the nsng training set, which is dominated by quasars, is limited to $i < 19.1$ \citep{richards:sdssqso}, which is reflected in the decline in the $r$-band number counts around $r \sim 19$. The logarithmic slopes of the galaxies and nsng are approximately 0.5 and 0.6 respectively for the training set. Stars, on the other hand, have a more heterogeneous selection scheme in the SDSS survey, with many subpopulations, which is reflected in the number count distribution of the stellar training set. 

We see clear evidence for reliable star-galaxy separation in Figure \ref{fig: mag counts}, where the nsng counts remain considerably lower than the star and galaxy counts until $r \lesssim 20$.  Given the spectroscopic limits of the SDSS, this figure suggests that the \texttt{photoPrimary} classifications are reliable to substantially fainter magnitudes than the spectroscopic sample. The galaxy counts approximately follow the logarithmic slope for around two magnitudes fainter than the training sample, and the nsng counts extend even fainter. The inflection points in the counts at $r \sim 20.5$ suggest that our classifications are reliable to this magnitude and fainter than this the tree becomes less reliable and increasingly assigns objects to nsng, which is one of the primary reasons we included this third classification in our algorithm. 

The SDSS photometric sample does extend fainter, the 95\% detection repeatability limits for point sources being $u < 22.0$, $g < 22.2$, $r < 22.2$, $i < 21.3$ and $z < 20.5$, which is seen in the number count distributions as the abrupt turnover at $r \gtrsim 22$. If, as the number counts suggest, our classifications are reliable to $r \lesssim 20$, the numbers of reliable classifications are 7,978,356, 832,993, and 13,945,621, totaling 22,756,970 for the galaxy, nsng, and star classes, respectively. We must emphasize, however, that the number of sources that will actually be scientifically useful will be lower, as these samples have not been cleaned. For example, before using these data, appropriate cuts should be made on either photometric errors, detection flags, or both, as described more fully on the SDSS project website.

\begin{figure}
\figurenum{8}
\plotone{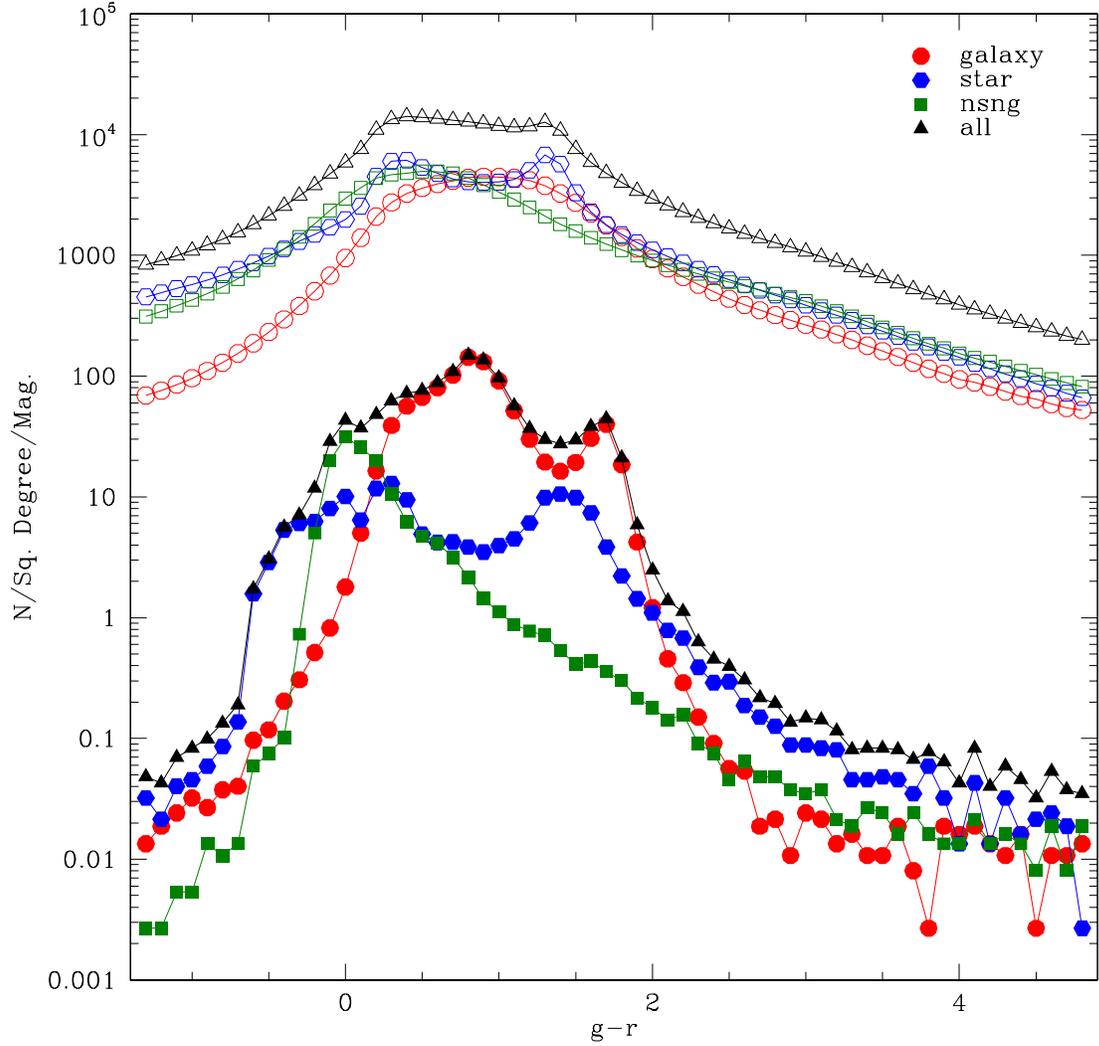}
\caption{As Figure \ref{fig: mag counts} but showing the objects in extinction-corrected model $g-r$ color. Bimodality in the galaxy and star color distributions is clearly seen in the training sets, but is washed out for galaxies in the full DR3 due to the higher mean redshift for the photometric sample. The nsngs are dominated by blue objects, as expected. The extrapolation in color space is considerably less than in magnitude space but is still present, e.g., the nsngs extend to much redder colors. The colors were used as the training set parameters. \label{fig: color counts}}
\end{figure}

Since we used colors, as opposed to magnitudes, to classify our sources, we also compared, as shown in  Figure~\ref{fig: color counts}, the $g - r$ distribution of training set sources with the \texttt{photoPrimary} classified sources. The extrapolation in $g - r$ color in going from the testing set to the classified sources is considerably less than the corresponding extrapolation in magnitude, although it is still present, particularly in the nsng class toward redder colors. As expected for the training set, galaxies and stars show a bimodal distribution in color and the nsngs are dominated by blue objects (e.g., UVX-selected quasars). This bimodality is washed out for galaxies in our classified sample, however, presumably due to the fact that the photometric sample has a higher mean redshift than the spectroscopic sample. We note that this difference in mean redshift will not affect the apparent magnitude number count distributions, as the redshift difference is too small to allow for significant evolution to occur between the two samples.

\begin{figure}
\figurenum{9}
\plotone{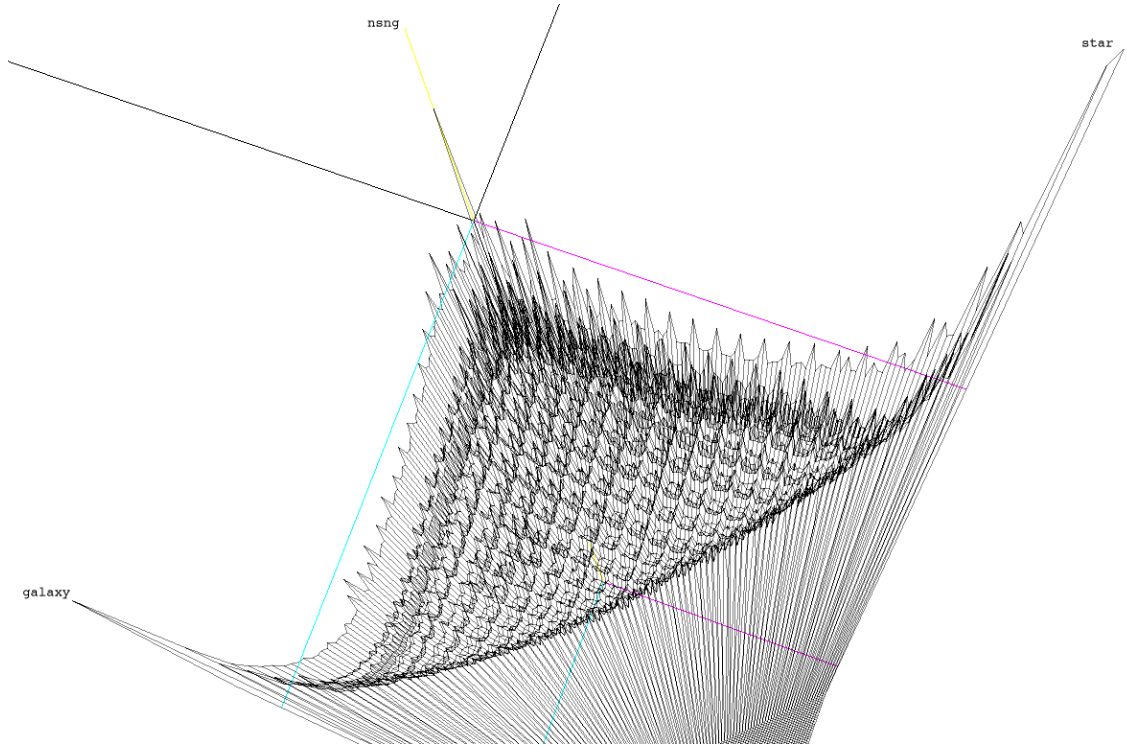}
\caption{Partiview visualization of the galaxy-nsng-star classification probabilities for the full \texttt{photoPrimary} catalog. The probabilities sum to 1 so $\probn$ is equal to 1 at the origin, with $\probg$ and $\probs$ ranging from 0 to 1 along their respective axes. The vertical axis shows the object counts---note the logarithmic scale. The upper horizontal axes are placed at a count of $10^5$ objects per bin. The clear separation of stars and galaxies is visible, as is the slightly less certain assignation of nsngs. Discretization in the probabilities can also be seen but is insignificant compared to the number of objects at high probability for each type. \label{fig: p(gns)}}
\end{figure}

In Figure~\ref{fig: p(gns)}, we present a visualization of the three classification probabilities for all objects in \texttt{photoPrimary}, demonstrating how the three values interact. Since the three probabilities must always sum to unity, their distribution appears as a triangle, with the height corresponding to the density of objects in that region of probability space. One can see that galaxies and stars are assigned relatively unambiguously, but that the tree is generally less certain when an object is neither a star nor a galaxy, as the base of the triangle rises noticeably before reaching $p_n = 1$. Discretization is also seen in this figure, but is insignificant compared to the numbers of each source type with high probability. The data used to produce the visualization are available to download with the catalog.

\subsection{Blind Testing} \label{Subsec: Results - Blind}

In the pseudo-blind test results described earlier (\S~\ref{Subsec: Results - Testing}), our decision tree provided robust classifications for data that was not used to train the decision tree, nor to characterize its performance. While this pseudo-blind test did provide a good approximation to a blind test, a true blind test is to compare the decision tree classifications to sources that have been assigned a spectroscopic identification from another survey. We have, therefore, matched the SDSS DR3 photometric data to the 2dFGRS and 2QZ surveys, by using coordinate position with a tolerance of 2 arcseconds. The 2dFGRS resulted in 50,191 matches and the 2QZ 10,259 matches. The 2dFGRS contains mainly galaxies and the 2QZ stars and quasars with a small number of mainly narrow emission-line galaxies.

\begin{figure}
\figurenum{10}
\plotone{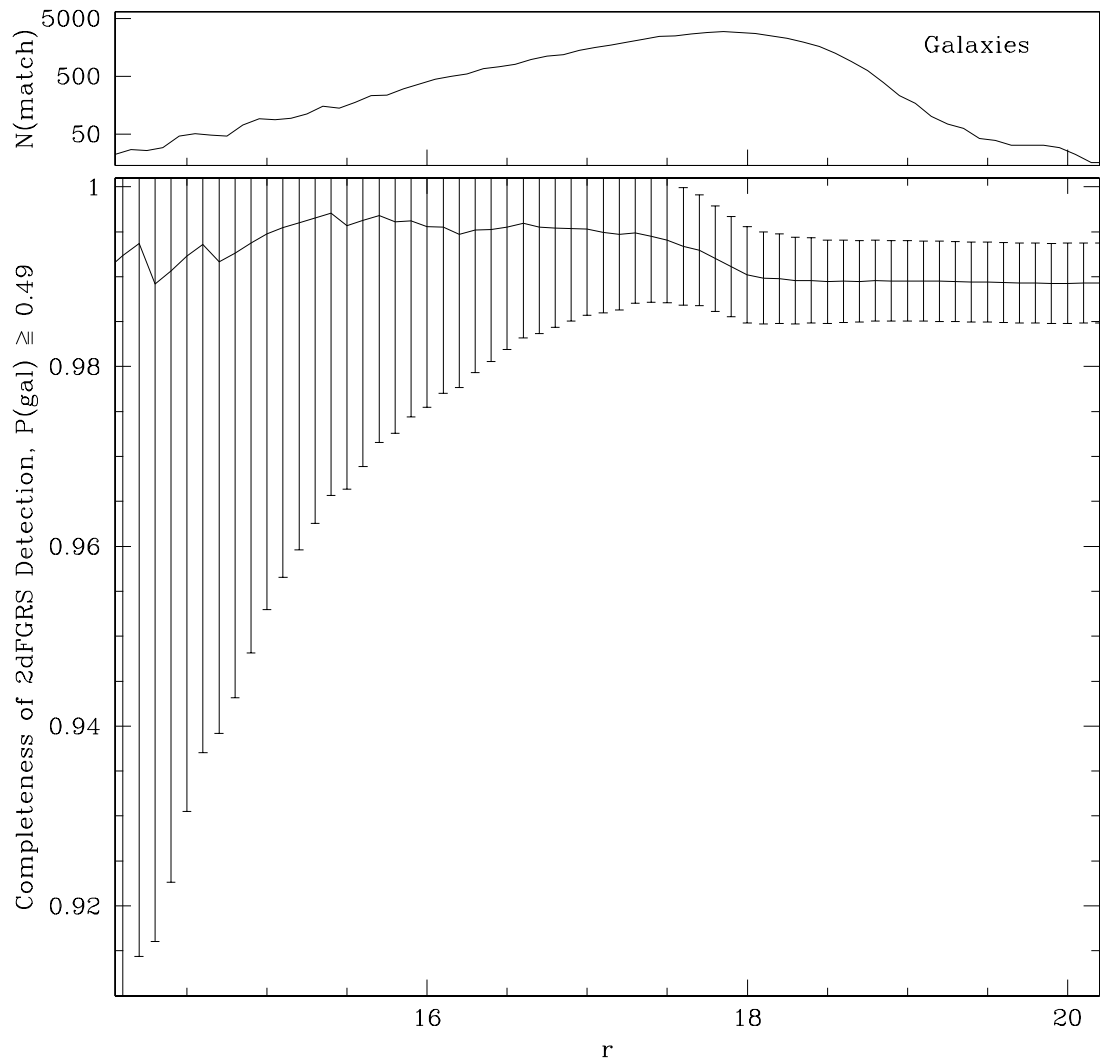}
\caption{Completeness as a function of magnitude for 2dFGRS galaxies. The upper panel shows the differential counts for matches with the SDSS DR3. The lower panel shows the completeness for the integrated counts. The error bars are Poisson. \label{fig: com eff galaxy}}
\end{figure}

\begin{figure}
\figurenum{11}
\plotone{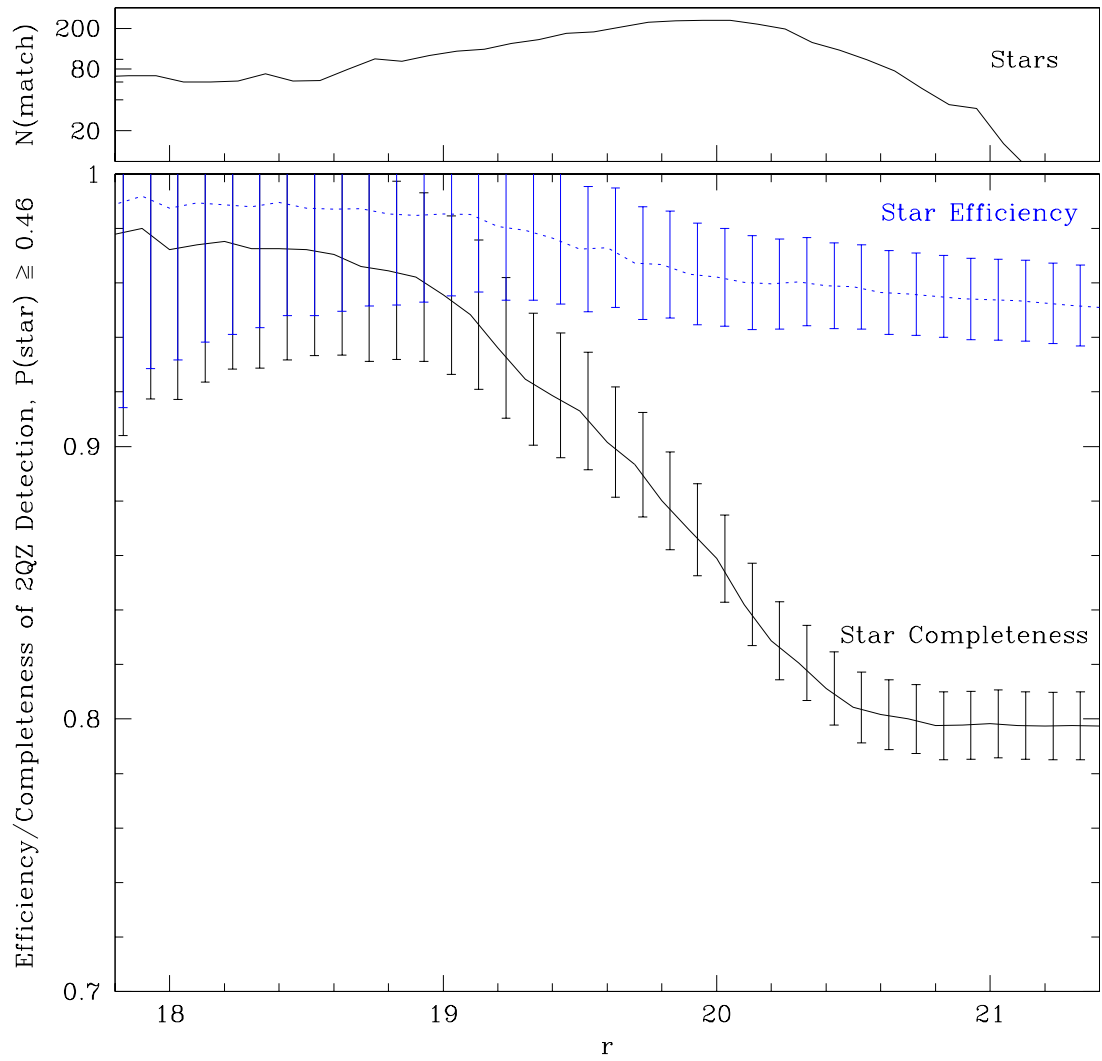}
\caption{As Figure \ref{fig: com eff galaxy} but showing completeness and efficiency for 2QZ stars. \label{fig: com eff star}}
\end{figure}

By applying the optimal cuts of $p_g \geq 0.49$, $p_n \geq 0.54$ and $p_s \geq 0.46$, as detailed in \S \ref{Subsec: Results - Testing}, we can characterize our classification accuracy in this blind test as a function of apparent magnitude. We show these results for the 2dFGRS galaxies in Figure~\ref{fig: com eff galaxy} for the 2QZ stars in  Figure~\ref{fig: com eff star}. For galaxies, the overall completeness of the 2dFGRS galaxies is 93.8\% (47,055 of 50,191), with the efficiency undefined as this is the only type of object. 
The completeness of the stars in the 2QZ match is rather low, at 79.7\% (4141 of 5193), but the efficiency is 95.4\% (4141 of 4359). Figure \ref{fig: com eff star} shows that the low completeness is due to stars at $r \gtrsim 19$. The low value may be due to the fact that the 2QZ is UVX-selected, which causes unusual stars that lie off the main-sequence, such as sub-dwarfs, halo stars, and hot young stars, to be preferentially targeted. 

\begin{figure}
\figurenum{12}
\plotone{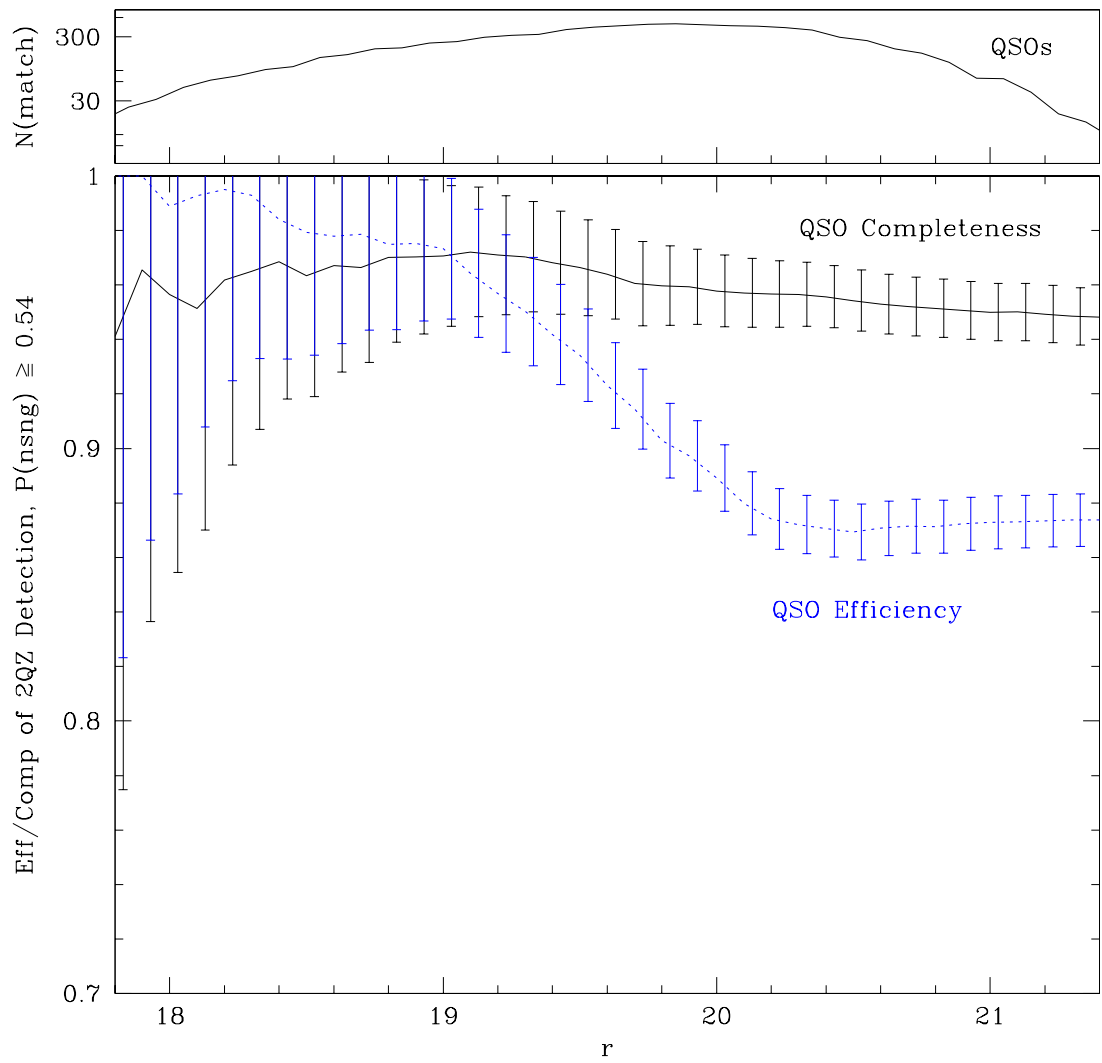}
\caption{As Figure \ref{fig: com eff star} but for 2QZ quasars. \label{fig: com eff quasar}}
\end{figure}

We also blind-tested the nsng classifications in a similar manner to the galaxy and star classes, but in this case by using quasars from the 2QZ with the same matching criteria as the one we used for stars from the 2QZ. As shown in Figure~\ref{fig: com eff quasar}, our decision tree does an excellent job of assigning quasars to the nsng class, with a completeness of 94.7\% (8278 of 8739) and an efficiency of 87.4\% (8278 of 9471) for the type \textit{11} best IDs and with a completeness of 94.4\% and an efficiency of 85\% for the \textit{12}, \textit{21} and \textit{22} next-best IDs. The efficiency drops off from around 98\% fainter than the SDSS spectroscopic limit of $i < 19.1$, and is about 87\% at $r \sim 20.5$ when the counts reach their limit in the 2QZ. This result is further vindication of adding the third class to our algorithm, as these sources are clearly neither stars nor galaxies, and our algorithm identifies them appropriately.  Overall, these blind tests support our assertion, based on the number count distributions, that our classifications remain reliable to $r \sim 20$.

\section{Discussion} \label{Sec: Discussion}

\subsection{Star-Galaxy Separation and Extrapolation} \label{Subsec: Discussion - star-galaxy}

The axis-parallel decision tree algorithm we have adopted in this work is an example of supervised machine learning. One of the primary criticisms of supervised techniques is their difficulty in extrapolating past the limits of the actual data used in their construction. Supervised learning algorithms can be viewed as a mapping between the training parameter space and a classification, and are not necessarily a representation of anything physical (e.g., a decision tree might correctly classify stars, but it would not necessarily reproduce the Hertzsprung-Russell diagram). Ideally, the training data should sample, perhaps sparsely, the entire parameter space occupied by the data to be classified.  In the case of astronomical datasets, such as the star-galaxy separation problem being studied in this paper, this is often impossible. We do not have a sufficiently large and faint spectroscopic survey to adequately sample either the magnitude or color spaces spanned by the photometric data from the SDSS survey. 

Yet we see that our classifications do reliably extrapolate past the training data. What is most likely driving this successful extrapolation is that the source colors do not change dramatically prior to $r \sim 20$. As a result, the existing training data is sufficient to classify sources in this area of parameter space. At these fainter magnitudes, the algorithm increasingly assigns sources to the nsng class, as the distinction between stars and galaxies is breaking down with decreasing signal-to-noise. We can characterize this effect from the classifications in the final catalog, where the fraction of nsng sources in the full catalog rises from approximately 1.3\% for $r \lesssim 18$ to approximately 3.7\% for $r \lesssim 20$. This observation, together with the obvious examples of sources that are neither stars nor galaxies, strengthens the case for introducing a third class into the historical star-galaxy separation problem. 

To aid in the utilization of our classification catalog, we have made available classifications for all sources in the SDSS DR3, without regard for photometric errors, or object detection flags. As a result, anyone wanting to use our catalog must subsequently apply any appropriate flag or photometric error restrictions as appropriate for their science---just as they would when using the SDSS databases directly. We feel that this is an important step forward in the realm of astronomical source classifications. While we are certainly cognizant that a large fraction of our source classifications are not highly reliable, we feel it is important in the new era of large surveys to provide an astronomer with the greatest possible freedom in selecting their own samples. If we released only those sources that pass a potentially opaque classification cut, we force all users of our catalog to deal with an additional, implicit selection effect when creating their own subsamples from our classification catalog.

\subsection{Future Work} \label{Subsec: Discussion - future}

In this paper, our supervised learning algorithm utilized photometry and spectra from the Sloan Digital Sky Survey. While the SDSS is an incredible resource for this sort of investigation, it is limited to the optical and near-infrared wavelengths, and, consequently, physical processes outside these wavelengths that might distinguish objects otherwise indistinguishable in the optical will not be seen. Given the growth in survey astronomy, a large number of SDSS objects have been observed in surveys at other wavelengths, and thus an important next-step is to include these in the training set. This extension has two stages: 1) match the SDSS photometry to spectroscopic targets from other surveys and train on the SDSS photometry with these augmented training data, and 2) add the photometry from other surveys as additional training parameters. The latter is a substantial task as the cumulative number of photometric objects in surveys such as GALEX \citep{martin:galex}, 2MASS \citep{skrutskie:2mass}, SWIRE \citep{lonsdale:swire} and ROSAT \citep{voges:rosat} number in the millions. We are in the process of matching the SDSS to these surveys, and others that may become available such as UKIDSS\footnote{\url{http://www.ukidss.org}}, to improve the classification performance achieved here. Of course this technique also extends to subsequent data releases from the SDSS itself.

We also would like to investigate alternative mechanisms for selecting SDSS photometric parameters for training, including the use of techniques such as forward-selection, backward-elimination, or a hybrid of the two, in which training parameters are respectively added or removed according to their effect on the quality of the classification. This approach is adopted by \citet{bazell:features}, who uses this technique for morphological galaxy classification. The training parameters could utilize magnitudes rather than colors, and also morphological parameters such as the $\texttt{probPSF}$ measure. In this paper, we have only used the colors of objects through four different matched apertures, as this work is our first major effort in large-scale machine classification of astronomical datasets. In the future, we anticipate performing a more detailed exploration of different classification parameters, especially as we incorporate additional datasets at other wavelengths into our algorithms. 

One training parameter that was found to not be useful in the current framework is the error on an object attribute, in agreement with previous analyses \citep[see, e.g.,][]{ball:ann}. Fundamentally, this is because, without additional information, the decision tree cannot differentiate between an error and another parameter. Coupled with the different dynamic ranges of the two types of attributes, this effect essentially trivializes the error attributes.  Since irrelevant attributes dilute the real information content in a training dataset, they reduce the performance of a decision tree. As a result, we did not include magnitude errors in our analysis. Intellectually, a better technique might be to include the error as part of the appropriate attribute when constructing a machine learning algorithm. Our decision tree implementation, however, does not provide this capability, and we therefore defer this concept to a future paper.

A further decision tree algorithm, supported in the Data-to-Knowledge framework but not yet extensively investigated, is \textit{boosting}, in which the differences between classes are emphasized. This method has been shown to improve results by many workers in the data mining field and is described further in \citet{hastie:learning}. Besides decision trees,  D2K supports numerous other algorithms, including na\"{i}ve Bayes, artificial neural networks, support vector machines, k-means clustering, and instance-based learning. The latter is of particular interest because, although it has shown promising results, its full potential has not been realized, due to the computational requirements of calculating the classifications for the full application set. 

In this method, the distance between each test or application set object (each instance) and the nearest $n$ neighbors in the training set is measured and a weighted mean is taken to give the new object its assigned type. Thus the training consists of specifying $n$, the weighting, and the distance measure, typically of the form $1 / d^x$ in each dimension where $d$ is the distance and $x=2$ would be the Euclidean distance. Hence there are a large number of distance calculations involved. Typically the classification of an object takes of the order of a second on a desktop machine, rendering the process computationally intractable for the 143 million objects in the SDSS DR3. The process is, however, embarrassingly parallel, so is amenable to a significant speedup by using supercomputing resources. In which case, this technique could potentially be used to not only classify sources, but to characterize the energy generation mechanisms responsible for a source's luminosity. We anticipate applying this technique, for example, to characterize the fraction of every source's luminosity in terms of the ratio between the thermonuclear fusion powering stars and the accretion powering quasars.

In many applications it has been found that a \textit{mixture of experts}, that is, the combination of results from more than one learning method, is best, in an analogous way to the improvement seen here through the use of bagging. A simple way to combine results is to take a weighted sum of outputs from the different methods. Preliminary tests have shown that instance-based with $n=20$ and $x=6$ in a 0.7:0.3 combination with a decision tree gives improved results. Of course, one must be able to run a full instance-based classification to explore the potential of combining it with the other methods, which heretofore has been impossible due to the necessary computing resources. Another promising future method is semi-supervised classification \citep{bazell:classdisc}, in which the algorithm is given some predefined classes but is able to discover further classes in the data for itself. This combines prior knowledge from spectra with the large number of objects for which there is available photometry and may perform better where very few spectra are available.

Finally, in this work we limited our analysis to three classes as our main interest was robust star-galaxy separation. Given the diversity of both stars and galaxies, and the ignorance of our approach to non-stellar luminosity, we could clearly benefit from adding additional classes to our existing decision tree work \citep[see, e.g.,][]{suchkov:dt}. As it stands, the nsng class is dominated by quasars at bright magnitudes, and could, therefore, serve as the basis for a quasar class that we expect would provide reasonable quasar classifications for relatively bright sources with clean photometry. Ideally, this approach could be extended to include other classes for different types of stars and galaxies, for example, white dwarfs, hot stars, cool dwarfs, starbursts, and active galaxies.

\section{Conclusions} \label{Sec: Conclusions}

In this paper we have classified all of the 142,705,734 non-repeat (primary) objects in the Third Data Release of the Sloan Digital Sky Survey (SDSS DR3). The classifications were determined by using machine learning, the algorithm of choice being the axis-parallel decision tree. The algorithm was trained on 477,068 objects for which spectroscopic classifications were available within DR3. This training data consists of 361,490 galaxies, 62,333 stars, 49,545 quasars, and 3,700 unknown objects. Collaboration with domain experts at the National Center for Supercomputing Applications (NCSA) and the use of the general machine learning environment Data-to-Knowledge combined with supercomputing resources enabled extensive investigation of the decision tree parameter space and the associated datasets. To our knowledge, this level of investigation has not previously been published in the astronomy literature.

This is the first public release of objects classified in this way for the whole survey. The objects are classified as either galaxy, star or nsng, and we provide an associated probability for each class. These three probabilities always sum to unity for every object classified. By merely assigning the classification with the highest probability, we find that our full classification sample contains 38,022,597 galaxies, 57,369,754 stars and 47,026,955 nsng in the magnitude range $0 < r < 40$, with 589 outside this range and 285,839 (0.2\%) ambiguous due to equal-highest probabilities for object type. The catalog is available for download at \url{http://quasar.astro.uiuc.edu/rml}. 

A major issue with this method of classification is the inevitable extrapolation from the spectroscopic regime. We investigate this by examining the apparent magnitude number counts for all sources and by performing several pseudo-blind or fully blind tests by using sources spectroscopically identified in the SDSS, 2dF Galaxy Redshift, and 2dF QSO Redshift (2QZ) surveys. We find that for a sample giving the optimal completeness and efficiency, the completeness values are 98.9\%, 91.6\%, and 88.5\% for galaxies, stars, and nsngs in the SDSS, and 94.7\% for quasars, which are classified as nsng, in the 2QZ. The corresponding efficiencies are 98.3\%, 93.5\%, 94.5\%, and 87.4\%. The number count distributions suggest that the classifications are reliable for $r \lesssim 20$, giving 7,978,356 galaxies, 832,993 nsng, and 13,945,621 stars. As we have not applied any restrictions to the classification catalog, such as limiting by photometric error or detection flags, the number of reliable sources will be lower; however, we provide these classifications to eliminate any opaque selection effects from our classification catalog, which simplifies the task of using these classifications to supplement additional analyses, such as as defining a target sample for follow-up spectroscopy.

The assignment of probabilities to each object allows one to investigate the completeness and efficiency of the classifications as a function of these probabilities. While ratios of the probabilities were investigated, we find that the samples with an optimal completeness and efficiency are yielded by the simple cuts $p_g \geq 0.49$, $p_n \geq 0.54$ and $p_s \geq 0.46$. Other values close to 0.5 also give very similar results. The full data describing the completeness and efficiency as a function of these three threshold probabilities is made available with the catalog.

We feel that the application of machine learning algorithms to large, astronomical surveys is a rich area of research. We are currently augmenting our training data with both additional wavelengths and fainter spectroscopic identifications. To improve our classification results, we are also performing additional tests to determine the optimal parameters for decision trees built from these data. Given the efficacy of our approach, we plan to increase the number of training classes used in our analysis, first by using just the SDSS, and later by using additional photometric and spectroscopic data. Finally, we are also exploring the efficacy of more powerful algorithms, such as instance-based classification, to tackle the fundamental goal of characterizing sources by the fraction of their energy that is derived from fusion and accretion. These results will presented in subsequent papers in this series.

\acknowledgments

NMB, RJB and ADM acknowledge support from NASA through grants NAG5-12578 and NAG5-12580, Microsoft Research, and from the NSF PACI Project. The authors made extensive use of the storage and computing facilities at the National Center for Supercomputing Applications and thank the technical staff for their assistance in enabling this work.

We thank Ani Thakar and Jan Van den Berg for help with obtaining a copy of the SDSS DR3 databases and clarifying the number of objects, and Stuart Levy for help with Partiview.

Funding for the SDSS and SDSS-II has been provided by the Alfred P. Sloan Foundation, the Participating Institutions, the National Science Foundation, the U.S. Department of Energy, the National Aeronautics and Space Administration, the Japanese Monbukagakusho, the Max Planck Society, and the Higher Education Funding Council for England. The SDSS Web Site is http://www.sdss.org/.

The SDSS is managed by the Astrophysical Research Consortium for the Participating Institutions. The Participating Institutions are the American Museum of Natural History, Astrophysical Institute Potsdam, University of Basel, Cambridge University, Case Western Reserve University, University of Chicago, Drexel University, Fermilab, the Institute for Advanced Study, the Japan Participation Group, Johns Hopkins University, the Joint Institute for Nuclear Astrophysics, the Kavli Institute for Particle Astrophysics and Cosmology, the Korean Scientist Group, the Chinese Academy of Sciences (LAMOST), Los Alamos National Laboratory, the Max-Planck-Institute for Astronomy (MPA), the Max-Planck-Institute for Astrophysics (MPIA), New Mexico State University, Ohio State University, University of Pittsburgh, University of Portsmouth, Princeton University, the United States Naval Observatory, and the University of Washington.

Data To Knowledge (D2K) software, D2K modules, and/or D2K itineraries, used by us, was developed at the National Center for Supercomputing Applications (NCSA) at the University of Illinois at Urbana-Champaign. 

This research has made use of NASA's Astrophysics Data System.

\clearpage

\begin{deluxetable}{cccc}
\tablenum{1}
\tablewidth{0pt}
\tablecaption{Mappings from the SDSS DR3 object spectroscopic type to decision tree target type. \label{table: mappings}}
\tablehead{\colhead{DR3 specClass} &\colhead{No. Objects} &\colhead{Target Type} &\colhead{No. Objects}}
\startdata
galaxy		&361,490		&galaxy		&361,490\\
star			&42,101 		&star		&62,333\\
star\_late	&20,232		&			&\\
unknown		&3,700		&nsng		&53,245\\
qso			&46,275		&			&\\
hiz\_qso	&	3,270		&			&\\
\enddata
\tablecomments{For the training, 13 objects (12 galaxies and 1 quasar) with clearly unphysical outlying values were removed.}
\end{deluxetable}

\clearpage

\begin{deluxetable}{ll}
\tabletypesize{\footnotesize}
\tablenum{2}
\tablecaption{Values of decision tree parameters tested during optimization. \label{table: dt params}}
\tablehead{\colhead{Decision Tree Parameter} &\colhead{Values}}
\startdata
Minimum decomposition population &32768 16384 8192 4096 2048 1024 512 256 128 64 32 16 8 4 2 1\\
Maximum tree depth &1 2 3 4 5 6 7 8 9 10 11 12 13 14 15 16 17 18 19 20\\
Minimum error reduction &999999.0 100000.0 10000.0 1000.0 100.0 10.0 1.0 0.1 0.01 0.001 0.0001 0.00001 0.000001 0.0\\
One half split &false true\\
Midpoint split &false true\\
Mean split &false true\\
Median split &false true\\
Number of repetitions &1 2 5 10\\
Fraction train examples &0.01 0.1 0.3 0.5 0.7 0.8 0.9 0.99\\
Number of bagging models &1 10 20 50\\ 
Bagging fraction &0.001 0.01 0.1 0.3 0.5 0.7 0.8 0.9 0.99 0.999\\
Random seed subsamples &1 ... 32\\
Random seed bagging &1 ... 32\\
\enddata
\tablecomments{Not all combinations of these parameters were tested. See \S \ref{Subsec: Results - DT optimization}}
\end{deluxetable}

\clearpage

\begin{deluxetable}{lcccc}
\tablenum{3}
\tablecaption{Confusion matrix for the SDSS DR3 blind testing set. \label{table: confusion matrix}}
\tablehead{\multicolumn{2}{c}{Assigned} &\multicolumn{3}{c}{Target}\\ \multicolumn{2}{c}{} &\colhead{galaxy} &\colhead{nsng} &\colhead{star}}
\startdata
		&\textit{galaxy}	&71,748 &608		&735\\
Number	&\textit{nsng}	&397    &9,514	&309\\
		&\textit{star}		&345    &451		&11,305\\
		\\
		\hline
		\\
		&\textit{galaxy}	&75.2	&0.637  &0.770\\
Percentage    &\textit{nsng}	&0.416	&9.97	&0.324\\
		&\textit{star}		&0.362	&0.473	&11.8\\
\enddata
\tablecomments{From a decision tree trained and tested on non-overlapping samples from an 80\% random subsample of the training data and subsequently applied to the other 20\%. There are 95,412 objects. One object was given equal $\probg$ and $\probn$ and is thus excluded.}
\end{deluxetable}

\clearpage

\begin{deluxetable}{ccccccc}
\tablenum{4}
\tablecaption{Probability cuts for the samples which maximize completeness and efficiency. \label{table: f(p) cuts}}
\tablehead{\colhead{f(p)} &\colhead{Target} &\colhead{Best min f(p)} &\colhead{Best max f(p)} &\colhead{Completeness (\%)} &\colhead{Efficiency (\%)} &\colhead{Average (\%)}}
\startdata
$\probg$ &galaxy      &0.49   &1.00   &98.9  &98.3 &98.6\\
$\probn$ & nsng     &0.54   &1.00   &88.5  &94.5 &91.5\\
$\probs$ &star        &0.46   &1.00   &91.6  &93.5 &92.6\\
$\probgn$ &galaxy  &0.821  &4.82   &98.2  &97.4 &97.8\\
$\probgn$ & nsng &-4.70  &-0.989 &76.4  &86.2 &81.3\\
$\probgn$ &star    &-2.32  &2.06   &94.2  &35.0 &64.6\\
$\probgs$ &galaxy  &0.770  &4.93   &97.6  &97.5 &97.6\\
$\probgs$ & nsng &-2.38  &1.97   &95.3  &32.3 &63.8\\
$\probgs$ &star    &-4.32  &-0.711 &74.3  &83.4 &78.9\\
$\probns$ &galaxy  &-1.62  &1.85   &96.7  &92.6 &94.7\\
$\probns$ & nsng &2.19   &4.02   &56.5  &90.0 &73.3\\
$\probns$ &star    &-4.23  &3.15   &100.0 &7.74 &53.9\\
\enddata
\tablecomments{The targets were split into galaxy, nsng, star. The ideal classifier would give a value of 100\% in the right-hand column. The columns not shown, e.g., $\probg$ with target nsng, give worse results. The inverse ratios give the same samples as those shown.}
\end{deluxetable}

\clearpage

\clearpage

\clearpage

\clearpage

\clearpage

\clearpage

\clearpage

\clearpage

\clearpage

\clearpage

\clearpage

\clearpage

\end{document}